\documentclass[letter]{aa} 

\usepackage{graphicx}
\usepackage[varg]{txfonts}
\begin{document} 

   \title{Hot bottom burning and s-process nucleosynthesis in massive
   AGB stars at the beginning of the thermally-pulsing phase}

   \titlerunning{Early massive AGBs}

   \author{D. A. Garc\'{\i}a-Hern\'andez\inst{1,2} \and O. Zamora\inst{1,2} \and
   A. Yag\"ue\inst{3,1,2} \and S. Uttenthaler\inst{4} \and A. I. Karakas\inst{5} \and M.
   Lugaro\inst{6} \and P. Ventura\inst{3} \and D. L. Lambert\inst{7}}

   \authorrunning{Garc\'{\i}a-Hern\'andez et al.}

          
   \institute{Instituto de Astrof\'{\i}sica de Canarias, C/ Via L\'actea s/n, E$-$38200 La Laguna, Spain \email{agarcia@iac.es} 
	\and Departamento de Astrof\'{\i}sica, Universidad de La Laguna (ULL), E$-$38206 La Laguna, Spain
        \and INAF-Osservatorio Astronomico di Roma, via Frascati 33, I$-$00040 Monteporzio, Italy
	\and University of Vienna, Department of Astrophysics, T\"urkenschanzstra\ss e 17, 1180 Vienna, Austria
	\and Research School of Astronomy \& Astrophysics, Australian National University, Canberra, ACT 2611, Australia
        \and Monash Centre for Astrophysics (MoCA), Monash University, Victoria, Australia 
        \and W. J. McDonald Observatory. The University of Texas at Austin. 1 University Station, C1400. Austin, TX 78712$-$0259, USA
	}


\date{Received April xx, 2013; accepted xxxx xx, 2013}

 
\abstract
{We report the first spectroscopic identification of massive Galactic asymptotic
giant branch (AGB) stars at the beginning of the thermal pulse (TP) phase. These
stars are the most Li-rich massive AGBs found to date, super Li-rich AGBs with
log$\varepsilon$(Li)$\sim$3-4. The high Li overabundances are accompanied by
weak or no s-process element (i.e. Rb and Zr) enhancements. A comparison of our
observations with the most recent hot bottom burning (HBB) and s-process
nucleosynthesis models confirms that HBB is strongly activated during the first
TPs but the $^{22}$Ne neutron source needs many more TP and third dredge-up
episodes to produce enough Rb at the stellar surface. We also show that the
short-lived element Tc, usually used as an indicator of AGB genuineness, is not
detected in massive AGBs which is in agreement with the theoretical predictions
when the $^{22}$Ne neutron source dominates the s-process nucleosynthesis.}

\keywords{Stars: AGB and post-AGB -- Stars: abundances -- Stars:evolution --
Nuclear reactions, nucleosynthesis, abundances -- Stars: atmospheres -- Stars:
late-type}

   \maketitle
%

\section{Introduction}

Low- and intermediate-mass (0.8 $<$ M $<$ 8 M$_{\odot}$) stars end their lives
with a phase of strong mass loss and thermal pulses (TP) on the asymptotic giant
branch (AGB, Herwig 2005), and are one of the main contributors to the chemical
enrichment (e.g. C, N, Li, F, and s-process elements) of the interstellar
medium. After each TP the surface convective zone moves inwards, and may reach
internal layers contaminated by 3$\alpha$ nucleosynthesis, enriched in C. This
mixing episode is known as third dredge-up (TDU). After many TDUs the surface C
exceeds the O content, transforming originally O-rich stars into C-rich stars.
However, this applies only to AGB stars in the mass range from $\sim$1.5 to 3-4
M$_{\odot}$. More massive stars experience hot bottom burning (HBB,  Sackmann \&
Boothroyd 1992), i.e. proton-capture nucleosynthesis at the base of the outer
envelope that favours the conversion of C to N by the CN-cycle and reconversion
of the C-rich to an O-rich atmosphere. The HBB process is activated when the
temperature at the bottom of the envelope reaches 40 MK, the same required to
ignite the Cameron \& Fowler (1971) mechanism (Mazzitelli et al. 1999), by which
great amounts of Li can be produced in the surface regions. The detection of Li
overabundances in massive O-rich AGB stars both in our Galaxy
(Garc\'{\i}a-Hern\'andez et al. 2007) and in the Magellanic Clouds (MCs) (e.g.
Plez et al. 1993; Garc\'{\i}a-Hern\'andez et al. 2009) has confirmed the
activation of HBB in the more massive AGB stars.

Theoretical nucleosynthesis models also predict the presence of the elements
(e.g. Rb, Zr, and Sr) that can be made by $slow$ neutron captures (the $s$
process) and are dredged up to the stellar surface. According to the most recent
models, $^{13}$C($\alpha$,n)$^{16}$O is the preferred neutron source for
$\sim$1-3 M$_{\odot}$, while for more massive stars neutrons are mainly released
by $^{22}$Ne($\alpha$,n)$^{25}$Mg (e.g. Karakas 2010; van Raai et al. 2012;
Karakas et al. 2012). The $^{22}$Ne neutron source requires higher temperatures
and produces higher neutron densities than the $^{13}$C source. The relative
abundance of Rb to other nearby s-elements such as Zr is very sensitive to the
neutron density owing to the operation of branchings in the s-process path at
$^{85}$Kr and $^{86}$Rb (Lambert et al. 1995; Abia et al. 2001; van Raai et al.
2012). In this context, the [Rb/Zr] ratio is a powerful discriminant of the
operation of the $^{13}$C versus the $^{22}$Ne neutron source in AGB stars.

A small group of Galactic stars showing OH maser emission, known as {\rm OH/IR}
stars, have been identified as massive (4-8 M$_{\odot}$) O-rich HBB AGB stars
showing strong Rb overabundances coupled with weak Zr enhancements
(Garc\'{\i}a-Hern\'andez et al. 2006, 2007). This provides the first
observational evidence that $^{22}$Ne is indeed the dominant neutron source in
massive Rb-rich AGB stars. However, this study was intentionally biased towards
the redder and more extreme Galactic {\rm OH/IR} stars, which experience very
strong mass-loss rates (up to $\sim$10$^{-4}$$-$10$^{-5}$ M$_{\odot}$/yr) and
are expected to be at the end of the thermally-pulsing (TP) AGB phase. A very
advanced evolutionary AGB stage is also suggested by the large variations in the
measured Li abundances ($-$1.0 $\leq$ log$\varepsilon$(Li) $\leq$2.6; see van
Raai et al. 2012); the Li-rich stars in that sample display Li abundances larger
than solar, but significantly smaller than those found in a few super Li-rich
Galactic AGB stars (with log$\varepsilon$(Li) $>$ 3$-$4; e.g. Abia et al.
1999)\footnote{We note that a few super Li-rich AGBs were found in the MCs (e.g.
Plez et al. 1993) but these are less massive ($\sim$3-5 M$_{\odot}$) Rb-poor
HBB-AGBs where the $^{13}$C neutron source dominates the s-process
nucleosynthesis (see Garc\'{\i}a-Hern\'andez et al. 2006, 2009).}.

Interestingly, HBB models predict that massive AGB stars experience a super
Li-rich phase (log$\varepsilon$(Li) $\sim$ 4) at the beginning of the TP phase
(e.g. Mazzitelli et al. 1999; van Raai et al. 2012). Yet, to date, no super
Li-rich massive Galactic AGB stars have been unambiguously identified (see e.g.
Uttenthaler \& Lebzelter 2010; Uttenthaler et al. 2011). In this {\it Letter} we
report the first detections of super Li-rich massive AGB stars in our Galaxy.
The extreme Li overabundances found together with the lack of s-process element
enhancements are consistent with these stars being truly massive O-rich AGB
stars at the beginning of the TP phase.

\section{Sample, optical observations, and chemical abundance analysis}

Our sample is composed of the Galactic disc O-rich AGB stars RU Cyg, SV Cas, R
Cen, and RU Ari. Their AGB status is deduced from their relatively long
pulsation periods ($\sim$200-500 days), large amplitude variability, late
spectral type ($>$M5), and infrared excess (see Garc\'{\i}a-Hern\'andez et al.
2007). All stars show similar infrared excesses and IRAS colours (R Cen and RU
Ari are the bluest and reddest sources, respectively), but very different
circumstellar maser emission from O-based molecules such as SiO, H$_2$O, and OH.
In Table A.1 (in the Appendix) we list the four AGB stars together with their
Galactic coordinates, present-day variability periods\footnote{Present-day
periods have been determined by us from AAVSO (http://www.aavso.org) visual
light curves (Uttenthaler et al. 2011).}, and maser information. An evolutionary
sequence of increasing mass-loss from SiO, to H$_{2}$O and OH maser emission is
expected in these stars (Lewis 1989; Habing 1996). Thus, the {\rm OH/IR} star RU
Ari is presumably more evolved (and/or more massive) than the other three AGBs
for which no OH maser emission has been detected. Also, defining marks of the
three {\rm non-OH/IR} AGBs are their complex (e.g. double maxima) and changing
variability properties. The dominant period of the Mira-like star R Cen has
decreased from $\sim$500 days in 2000 (Hawkins et al. 2001) to the present value
of 251 days while the other two stars have increased their dominant periods from
$\sim$200-300 days to $\sim$450 days. It has been speculated that the changing
period is related to a recent TP in AGB stars (see e.g. Wood \& Zarro 1981;
Uttenthaler et al. 2011).

The optical observations of the stars RU Cyg, SV Cas, and RU Ari presented here
are part of our recent high-resolution (R$\sim$60,000) optical spectroscopic
survey of a large sample ($\sim$100) of relatively blue Galactic O-rich AGB
stars not previously studied. This survey has been carried out in several
observing runs from 2006 to 2008 by using the Tull (Tull et al. 1995) and SARG
spectrographs at the 2.7m Harlan J. Smith (HJS) Telescope and at the Telescopio
Nazionale Galileo, respectively; the optical spectra and the abundance analysis
for the whole sample will be presented in a future paper. In particular, the
optical spectra reported here were taken in the wavelength range
$\sim$3650$-$9500 \AA~(R$\sim$60,000 with the 1.2" slit and the grating E2) at
the HJS telescope. On the other hand, the unpublished high-resolution
(R$\sim$50,000) optical spectrum of R Cen was taken with the ESO-VLT UVES
spectrograph (with the 0.7" slit), covering the $\sim$3770$-$4900, 6670$-$8470,
and 8650$-$10500 \AA~spectral regions. The star R Cen was observed within the
ESO programme 65.L-0317(A) that was devoted to studying AGB stars in the
Galactic bulge (GB; see e.g. Uttenthaler et al. 2007). However, R Cen is a
low-latitude (b=$+$1.2$\,^{\circ}$) Galactic disc AGB star and displays a Li I
6708\AA~line that is much stronger than in the AGBs studied by Uttenthaler et
al. (2007). The exposure times varied between a few seconds (R Cen) and 3-5
minutes (RU Cyg and SV Cas) for the bluer AGBs and 20 minutes for the more
obscured star RU Ari. A signal-to-noise ratio (S/N) of $\geq$30$-$50 at 4200
\AA~and in excess of 100 at wavelengths longer than 5900 \AA~was achieved in the
blue AGBs. The only exception was the redder AGB star RU Ari for which a S/N
higher than 10 was achieved at 5924 \AA~(the Tc I line) and higher than 40 at
$\lambda$$>$6400 \AA. 

The two-dimensional optical {\it echelle} spectra were reduced to single-order
one-dimensional spectra using standard astronomical tasks. We are mainly
interested in the wavelength regions around the Li I 6708\AA~and Rb I
7800\AA~resonance lines and around the ZrO 6474\AA~bandhead. The Galactic O-rich
AGB stars analysed here are bright enough in the 4000$-$6000 \AA~region,
permitting us to search for the Tc absorption lines (e.g. at $\sim$4238, 4262,
4297, and 5924 \AA) for the first time. This is in strong contrast with the more
obscured ({\rm OH/IR}) massive Rb-rich AGB stars previously studied in our
Galaxy (Garc\'{\i}a-Hern\'andez et al. 2006, 2007) that are too faint at
wavelengths below 6000 \AA. Indeed, the blue AGBs in our sample do not display
the still unidentified molecular bands that are clearly present in obscured
massive Rb-rich AGB spectra (e.g. at $\sim$7400$-$7600 \AA; see
Garc\'{\i}a-Hern\'andez et al. 2009). These unidentified molecular bands are
also detected in the much redder ({\rm OH/IR})AGB star RU Ari. This
spectroscopic difference permit us to double-check the Zr abundance derived from
the ZrO molecular bands with that estimated from the lack of the Zr I lines in
the 7400-7600 \AA~region in the blue AGBs (see below). 

The chemical abundance analysis follows the procedure previouly used by us in
Galactic O-rich AGB stars (see Garc\'{\i}a-Hern\'andez et al. 2006, 2007 for
more details). In short, we used the latest version (v12.1.1) of the
TURBOSPECTRUM package (Alvarez \& Plez 1998; Plez 2012) with line-blanketed
model atmospheres and up-to-date molecular/atomic linelists to construct a grid
of synthetic spectra appropiate for cool O-rich AGB stars in the Galactic disc:
{\it T}$_{eff}$=2600-3500 K in steps of 100 K, FWHM=200-600 m\AA~in steps of 50
m\AA, log {\it g}=-0.5, $\xi$=3 kms$^{-1}$, and solar metallicity ([M/H]=0.0).

The observations were compared to the synthetic spectra in the regions
6455$-$6499, 6670$-$6730, and 7775$-$7835 \AA, covering the ZrO
6474\AA~bandhead, the Li I 6708\AA~line, and the Rb I 7800\AA~line (Figures A.1
and A.2). The best fit to the TiO bandheads and the pseudocontinuum around the
atomic lines provided the {\it T}$_{eff}$, which was found to be 3000$\pm$100 K
for all sample stars. The abundances (or upper limits) of Li, Rb, and Zr were
obtained by adjusting the Li I and Rb I lines and the ZrO molecular bands,
respectively. In the blue AGBs, we note that the Zr abundance derived from the
ZrO molecular bands is consistent with the lack of the Zr I lines in the
7400-7600 \AA~region. The spectroscopic {\it T}$_{eff}$ and abundances are given
in Table A.1, where we also list the abundance uncertainties estimated for each
star. The three bluer AGB stars show strong Li enhancements
(log$\varepsilon$(Li) $\geq$ 2) together with no (or small) Rb and Zr
overabundances (log$\varepsilon$(Rb, Zr)$\sim$2.6$\pm$0.3). However, the redder
AGB RU Ari displays a very high Rb content accompanied by no Li and no (or weak)
Zr (see Sect. 4).

Finally, we have searched for the presence of the strong Tc I lines at
$\sim$4238, 4262, and 4297 \AA~and the much weaker Tc I 5924\AA~line by
following the method described by Uttenthaler et al. (2011). We note that the
bluest AGB R Cen was not observed around the Tc I 5924\AA~line and the reddest
AGB star RU Ari is too faint around the stronger Tc I lines at $\sim$4200$-$4300
\AA. Our Tc search proved to be negative and all sample stars are non-Tc stars
(see an example in Fig. A.3). The non-detection of Tc is also suggested by
spectrum synthesis around the Tc I 5924\AA~line\footnote{The atomic parameters
were taken from Palmeri et al. (2005).} (see Fig. A.2). Our synthetic spectra
suggest that Tc is not detectable (log$\varepsilon$(Tc)$<$0.0) in our stars,
although the Tc I 5924\AA~line is not very sensitive to the Tc content.

\section{The super Li-rich massive AGB stars}

The stars SV Cas and R Cen are among the most Li-rich (super Li-rich with
log$\varepsilon$(Li) $>$ 3$-$4) O-rich (M-type) AGB stars found in our
Galaxy\footnote{We mote that the O-rich and Tc-poor (M-type) AGB star R Nor was
previously found to be extremely enriched in Li (with log$\varepsilon$(Li)=4.6)
by Uttenthaler et al. (2011). The star R Nor (with {\it T}$_{eff}$=3000$\pm$100
K) is a spectroscopic twin of R Cen. Our spectrum synthesis confirms that R Nor
is a super Li-rich (with log$\varepsilon$(Li)=4.0) AGB with no (or weak) Zr
(log$\varepsilon$(Zr)=2.6$\pm$0.2) enhancement. However, the R Nor spectrum does
not cover the Rb I lines at 7800 and 7947 \AA. We expect R Nor to be
non-enriched in Rb, being another example of super Li-rich massive AGBs at the
beginning of the TP phase (see text).}. The high Li overabundances observed in
these Galactic disc O-rich AGBs are most likely explained by Li-enrichment due
to HBB. The evolution of the Li abundance in HBB-AGB stars strongly depends on
several stellar parameters such as progenitor mass, metallicity, mass loss, and
convection model (e.g. Mazzitelli et al. 1999; van Raai et al. 2012). The
treatment of convection and of mass loss are the most important ingredients in
determining the duration of HBB and the variation of the surface chemistry
during the AGB phase. This particularly holds in the range of masses
experiencing HBB (Ventura \& D'Antona 2005a,b). Thus, we compare our
observations with up-to-date solar metallicity HBB model predictions with very
different prescriptions for mass loss and convection: i) models with the
Bl\"ocker (1995) recipe for mass loss and the full spectrum of turbulence
convective mixing (HBB-FST; e.g. Mazzitelli et al. 1999); and ii) models with
the Vassiliadis \& Wood (1993) (VW93) delayed mass loss and the mixing length
theory for convection (HBB-MLT; see discussion in Karakas et al. 2012).

The temporal variation of the surface Li abundance in up-to-date HBB-FST models
of 3.5 M$_{\odot}$ $\leq$ M $\leq$ 6 M$_{\odot}$ is displayed in Fig. 1 (left
panel). The new models reach higher Li values compared to those previously
published by Mazzitellli et al. (1999) in which only the 6 M$_{\odot}$ model
reached log$\varepsilon$(Li) $>$ 4. The corresponding Li predictions from the
most recent HBB-MLT models (Karakas et al. 2012) are very similar with the
exception that the 4 M$_{\odot}$ model does not become Li-rich (Karakas 2010).
Figure 1 (right panel) shows these predictions for a 5 M$_{\odot}$ model. There
are two obvious differences between the two sets of models. First, the Li minima
in the HBB-FST models are deeper than in the HBB-MLT models, probably because of
the different efficiency of the convection model. When the TP is on (or has just
extinguished), the HBB-FST models are cooler than the HBB-MLT models and this
stops Li production for a while, leaving room for Li destruction via proton
capture. Second, the evolution of the HBB-FST models (and consequently the
duration of the Li-rich phase) is shorter than their HBB-MLT counterparts. The
stronger HBB predicted by the FST model (Ventura \& D'Antona 2005a) leads to a
fast increase in the stellar luminosity. The use of the Bl\"ocker mass loss
recipe in the HBB-FST models further enhances the difference in the evolutionary
times of the two sets of models (Ventura \& D'Antona 2005b).

\begin{figure*}
\centering
\includegraphics[angle=0,scale=.35]{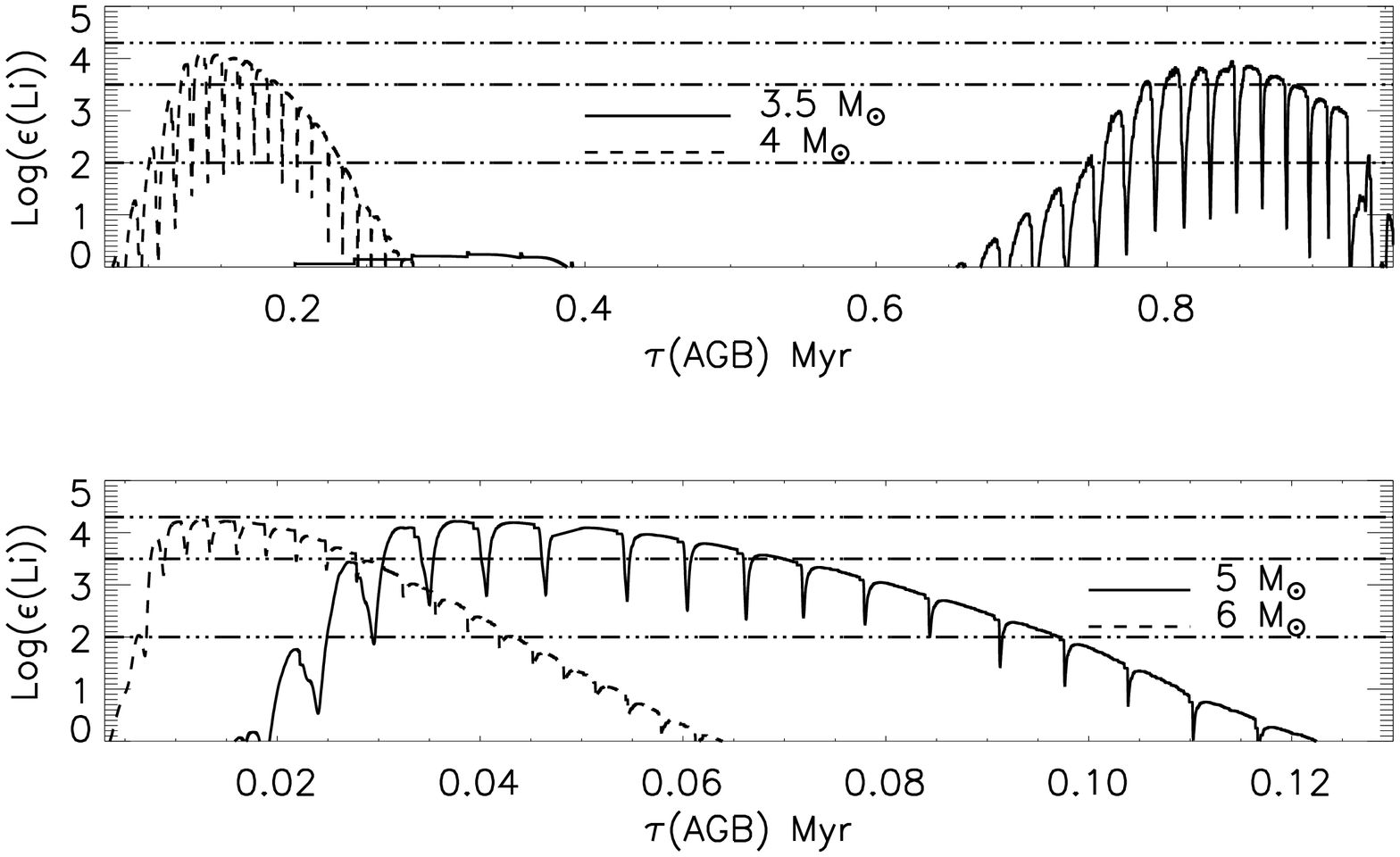}%
\includegraphics[angle=90,scale=.30]{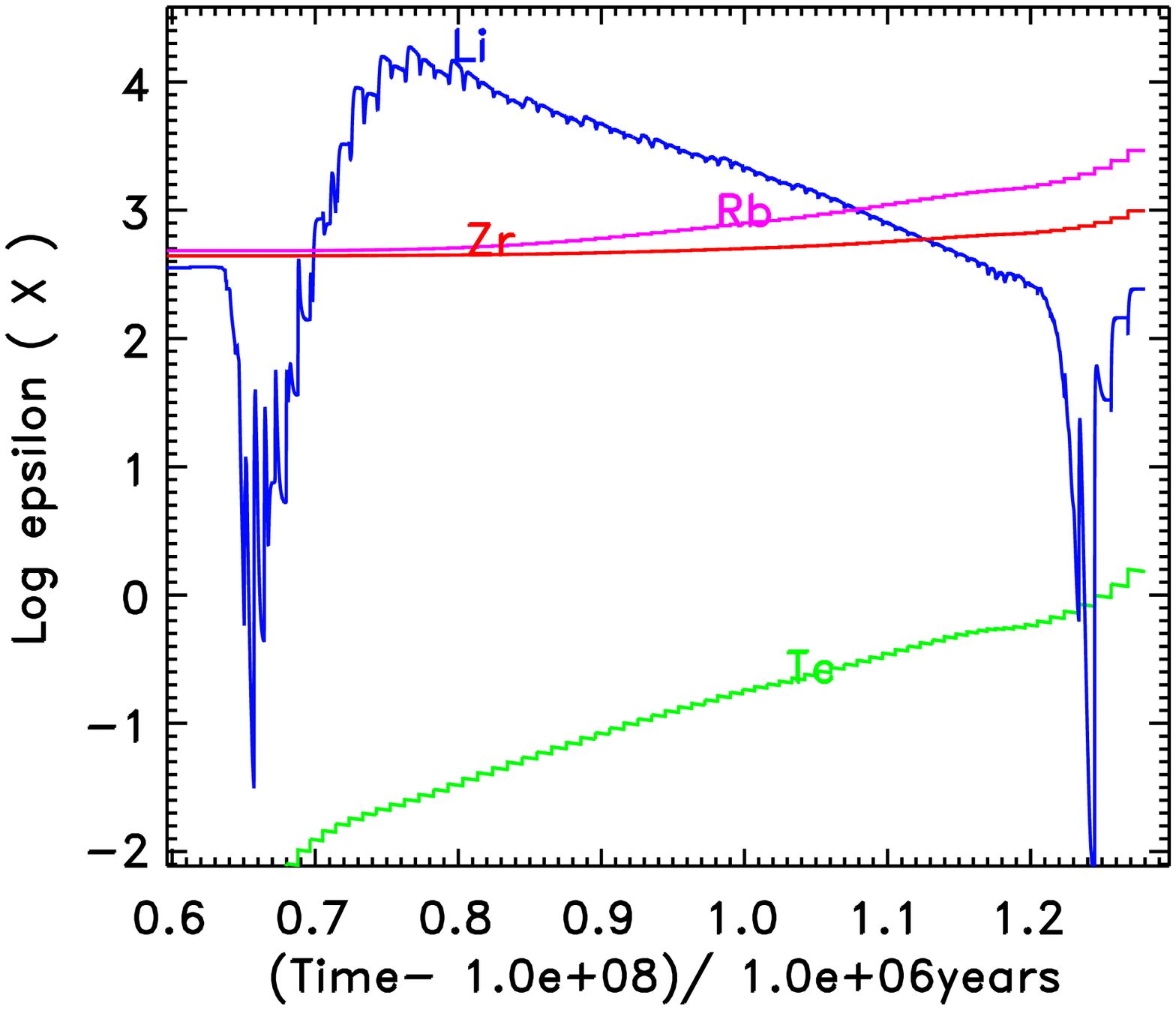}
\caption{Left panel: HBB-FST predictions of the Li abundance versus time from
the beginning of the AGB (in Myr). The horizontal lines represent the observed
values of log$\varepsilon$(Li)=2.0, 3.5, and 4.3. Right panel: HBB-MLT Li, Rb,
Zr, and Tc predictions for a 5 M$_{\odot}$ model with a delayed superwind. We
note that the Tc abundance is an upper limit because the trend of the half-life
of $^{99}$Tc - decreasing with temperature from terrestrial 0.22 Myr to $\sim$3
yr at the peak temperature of $\sim$350 MK - is not included in the model
(though we do not expect large differences, see Mathews et al. 1986).}
\label{Fig1}
\end{figure*}

Figure 1 shows that log$\varepsilon$(Li)$\sim$4 is reached during the first TPs.
In the HBB-FST scenario this occurs at the 3$^{rd}$, 5$^{th}$, and 11$^{th}$ TP
for the 6, 5, and 4 M$_{\odot}$ models, respectively; in the HBB-MLT framework
at the 7$^{th}$ and 14$^{th}$ TP for the 6 and 5 M$_{\odot}$ models,
respectively. The HBB models also predict 3$<$$^{12}$C/$^{13}$C$<$20 during the
super Li-rich phase, as expected. The HBB-FST models with mass below 4
M$_{\odot}$ (Fig.1, left panel) are also expected to become Li-rich but only in
the late AGB phase when the many TP and TDU episodes would favour an enhancement
of s-process elements like Rb (and Zr), which is not observed (see Sect. 4). In
short, we conclude that our observations of super Li-rich AGBs are consistent
with AGB stars of $\sim$4-6 M$_{\odot}$ at the beginning of the TP phase. The
situation may be more complex for the AGB star RU Cyg with
log$\varepsilon$(Li)=2. This star is probably in an inter-pulse period, on its
way to becoming a super Li-rich star, or has just been one. An early AGB stage
for these blue HBB-AGB stars is also suggested by other observational properties
such as the low changing periods, low infrared excess, and no OH maser emission
(Table A.1) and seems to be corroborated by the specific s-process
nucleosynthesis experienced by these stars (see below).

\section{The s-process in massive AGBs at the beginning of the thermally-pulsing
phase}

Remarkably, the three bluer AGBs display high Li overabundances accompanied by
very weak or no Rb and Zr enhancements (Table A.1) while the redder {\rm OH/IR}
AGB RU Ari shows a chemical abundance pattern (high Rb together with no Li and
weak Zr) that is typical of Rb-rich massive Galactic AGBs where the $^{22}$Ne
neutron source dominates the s-process nucleosynthesis (Garc\'{\i}a-Hern\'andez
et al. 2006, 2007). It is to be noted here that the [Rb/Zr] ratios observed in
Rb-rich massive {\rm OH/IR} AGBs such as RU Ari are much larger than predicted
by the s-process models (see Garc\'{\i}a-Hern\'andez et al. 2009). The authors
van Raai et al. (2012) discuss several possibilities (e.g. a failure of the
present model atmospheres which may give to Rb overestimations, condensation of
Zr to dust, etc.) that could explain this Rb/Zr mismatch. However, the situation
is very different in the super Li-rich massive AGB stars presented here. These
stars are very blue (little circumstellar dust, no OH maser emission) and are
not as extreme as the more evolved and Rb-rich AGBs. Thus, the present model
atmospheres and the derived abundances in the super Li-rich massive AGBs are
expected to be more reliable (see also Sect. 2).

The short-lived element Tc (usually used as an indicator of an intrinsic AGB
star, see below), however, is not detected (i.e. log$\varepsilon$(Tc)$<$0) in
all sample stars\footnote{We note that the detection threshold for Tc in AGB
stars is log$\varepsilon$(Tc)$\sim$0 (see Fig. 15 in Goriely \& Mowlavi 2000).}.
The lack of Tc can be explained by the operation of the $^{22}$Ne neutron source
in massive AGB stars. The s-process nucleosynthesis models that better reproduce
the large Rb overabundances observed in these stars are the delayed superwind
models by Karakas et al. (2012), where the production of s-process elements is
dominated by the $^{22}$Ne neutron source. Their solar metallicity 5 and 6
M$_{\odot}$ models show that the super Li-rich phase is reached at the beginning
of the TP phase, but it is not until much later (i.e. by the 33$^{rd}$ and
30$^{th}$ TP, respectively) that the s-process abundances (e.g. Rb) start to
rise. Interestingly, log$\varepsilon$(Tc)$<$0 during all AGB evolution. This is
shown in Fig. 1 (right panel) where we display the Rb, Zr, and Tc predictions
for a 5 M$_{\odot}$ model with a delayed superwind (Karakas et al. 2012). The
activation of the $^{13}$C neutron source predicts a much higher and faster Tc
production (also Zr production but at a much slower rate), which is not
observed. Models that include a $^{13}$C-rich region (the $^{13}$C pocket) in
the He intershell of massive AGB stars (see Karakas et al. 2012 and Lugaro et
al. 2012) predict a fast increase of the Tc abundance at the beginning of the TP
phase or during the super Li-rich phase; e.g. log$\varepsilon$(Tc)$>$0 by the
13$^{th}$ TP in a 5 M$_{\odot}$ model (see Fig. A.4). The production of Tc
depends on the neutron exposure experienced by the intershell material,
specifically, if its value is high enough to allow bypassing the first s-process
peak at Sr, Y, Zr to reach Tc. The neutron exposure from the $^{22}$Ne source
builds up in time, pulse after pulse, which is why the Tc abundance linearly
increases with the pulse number. On the other hand, the neutron exposure in the
$^{13}$C pocket is larger and the same in each $^{13}$C pocket, which means that
Tc is already produced in the very first $^{13}$C pocket. Thus, we conclude that
the lack of Tc in massive AGB stars is the consequence of $^{22}$Ne being the
dominant neutron source at the s-process site, as predicted by Goriely \& Siess
(2004).

The detection of Tc in s-process enriched (MS-, S-, SC-, and C-type) AGB stars
has been widely used as an indicator of AGB genuineness (``intrinsic'' stars),
while its non-detection is indicative of binary systems (``extrinsic'' stars)
(see e.g. Van Eck \& Jorissen 1999; Abia et al. 2002). This is correct if the
$^{13}$C neutron source dominates the s-process nucleosynthesis, as in low-mass
($\sim$1-3 M$_{\odot}$) AGB stars. However, we have shown here that the use of
Tc as an indicator of AGB genuineness is not valid for more massive AGB stars
for which the $^{22}$Ne is the dominant neutron source. Notably, many O-rich
(M-type) Mira-like stars without Tc, among them our sample star R Cen, were
found by Little et al. (1987). We propose that an important fraction of these
Tc-poor stars may be super Li-rich massive AGBs at the beginning of the TP
phase. Indeed, Little et al. (1987) found that $^{12}$C/$^{13}$C$\sim$8 for the
three Miras without Tc in their sample, which is very near to the equilibrium
value expected for strong HBB AGB stars. 

\begin{acknowledgements}
This work is partially based on observations made at ESO, 65.L-0317(A). D.A.G.H.
acknowledges support for this work provided by the Spanish Ministry of Economy
and Competitiveness under grants AYA$-$2011$-$27754 and AYA$-$2011$-$29060. S.U.
acknowledges support from the Austrian Science Fund (FWF) under project
P~22911-N16. A.I.K. is grateful for the support of the NCI National Facility at
the ANU and the ARC for support through a Future Fellowship (FT110100475). M.L.
thanks the ARC for support through a Future Fellowship (FT100100305) and Monash
University for support through a Monash Research Fellowship. D.L.L. wishes to
thank the Robert A. Welch Foundation of Houston, Texas, for support through
grant F-634.
\end{acknowledgements}

\Online
\begin{appendix}
\section{Table A.1 and Figures A.1$-$A.4}

\begin{table*}
\tiny
\caption{\label{t1} The sample of early massive AGBs: spectroscopic {\it T}$_{eff}$
and derived Li, Rb, and Zr abundances\tablefootmark{a}.}
\centering
\begin{tabular}{lccccccccc}
\hline\hline
Star   &  Gal. Coor. & {\it T}$_{eff}$   &  log$\varepsilon$(Li)  & log$\varepsilon$(Rb) &
log$\varepsilon$(Zr) &  Tc &  Period\tablefootmark{b} & Masers & \\

 & (l, b) & K & log{\it N}(Li) $+$ 12 & log{\it N}(Rb) $+$ 12& log{\it N}(Zr) $+$ 12&  & (days) & SiO/H$_{2}$O/OH & Ref.\tablefootmark{c} \\
\hline
RU Cyg & (97.37,$+$1.20)   & 3000 & 2.0$\pm$0.5  & $<$2.6$\pm$0.3 & 2.6$\pm$0.2	   & No	     & 442     & No/No/No            & 1/1/1              \\
SV Cas & (111.83,-9.04)    & 3000 & 3.5$\pm$0.5  & 2.6$\pm$0.3    & 2.7$\pm$0.2	   & No	     & 456     & No/$\dots$/$\dots$  & 2/$\dots$/$\dots$ \\
R Cen\tablefootmark{d}  & (313.42, $+$1.21) & 3000 & 4.3$\pm$0.5  & 2.6$\pm$0.3    & $\dots$        & No      & 251     & No/Yes/No           & 1/1/1              \\
RU Ari & (161.47,-41.90)   & 3000 & No           & 4.5$\pm$0.7    & $<$2.6$\pm$0.2 & No      & 357     & Yes/Yes/Yes         & 3/1/1              \\
\hline
\end{tabular}
\tablefoot{
\\
\tablefoottext{a}{The abundance uncertainties represent the formal
errors due to the sensitivity of the derived abundances to slight changes in the
model atmosphere parameters ($\Delta${\it T}$_{eff}$=$\pm$100 K, $\Delta$[M/H]=$\pm$0.3,
$\Delta$$\xi$=$\pm$1 kms$^{-1}$, $\Delta$log {\it g}=$+$0.5, $\Delta$FWHM=50 m\AA) for
each star.}
\tablefoottext{b}{Present-day pulsation periods derived by us from AAVSO data.}
\tablefoottext{c}{References for the SiO, H$_{2}$O, and OH maser observations.}
\tablefoottext{d}{R Cen was not observed around the ZrO 6474\AA~bandhead, but its
spectrum around the weaker (and less sensitive to the Zr content) ZrO
bands near 6925 \AA~is identical to the other stars.}}
\tablebib{(1) Benson et al. (1990); (2) Spencer et al. (1981); (3) Deguchi et al. (2012).}
\end{table*}

\begin{figure*}
\centering
\includegraphics[angle=0,scale=.35]{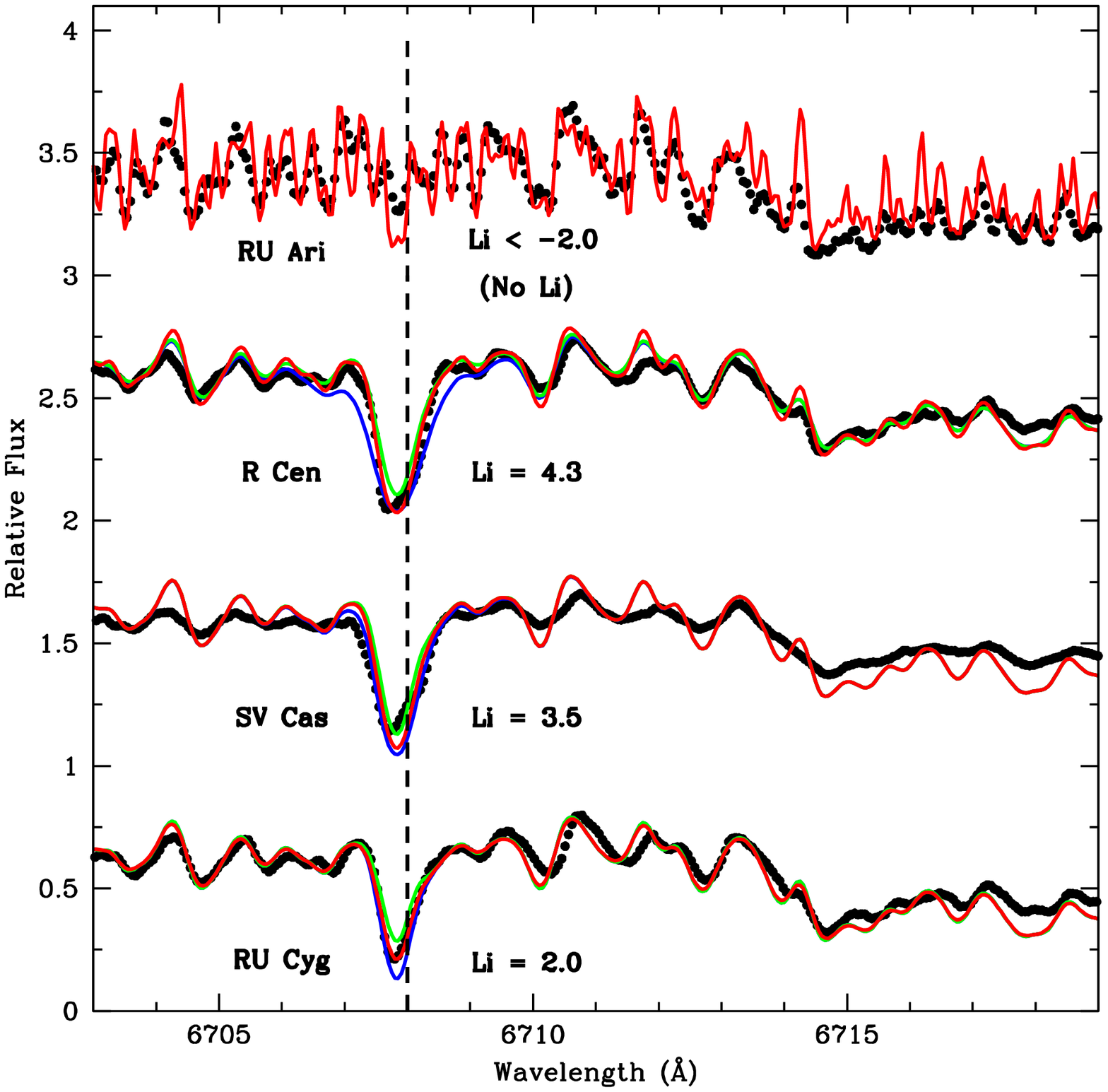}%
\includegraphics[angle=0,scale=.35]{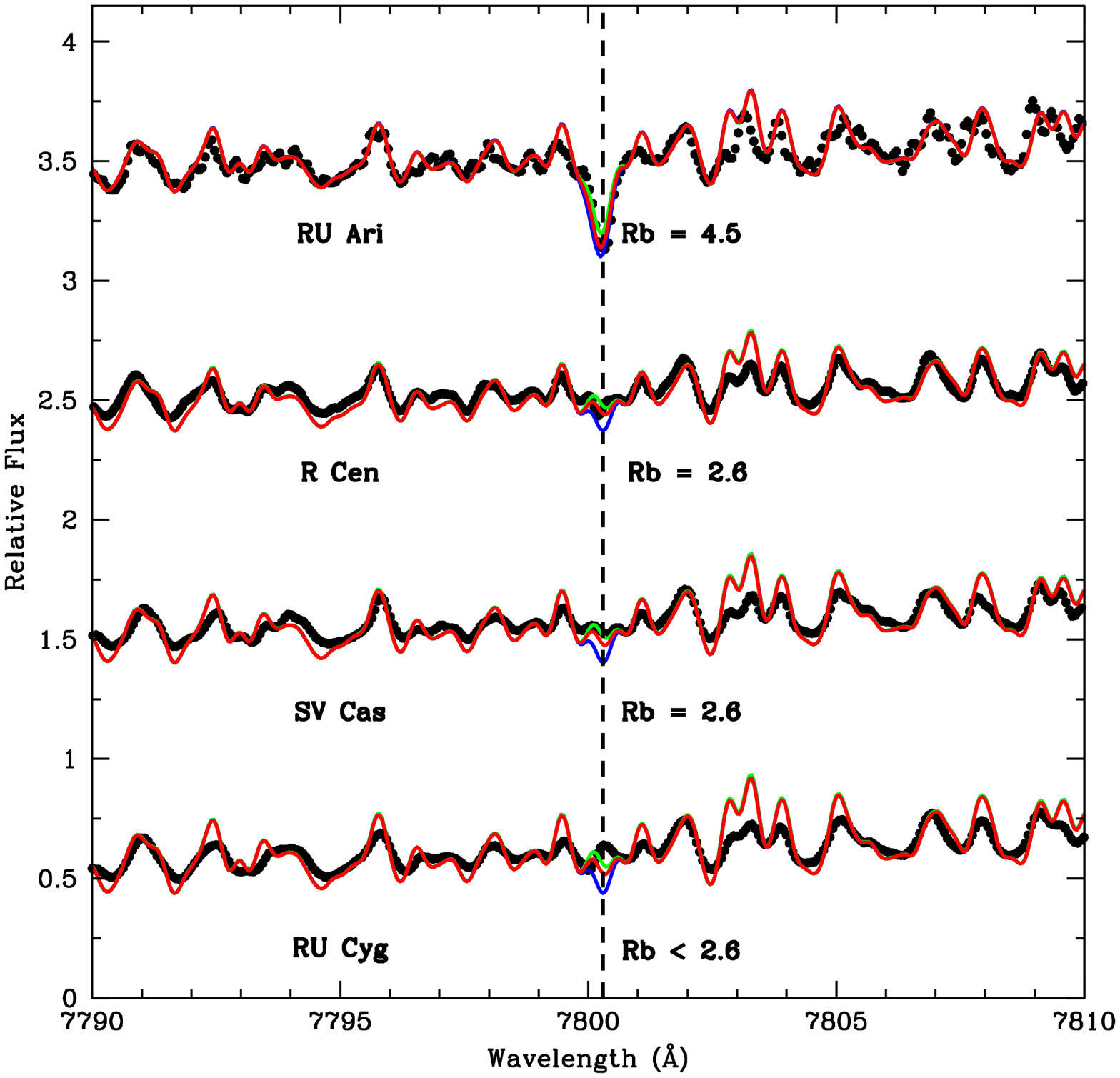}
\caption{High-resolution optical spectra (in black) and best model fits (in red)
in the Li I 6708\AA~region (left panel) and Rb I 7800\AA~region (right panel)
for the AGB stars RU Ari, R Cen, SV Cas, and RU Cyg. The derived Li and Rb
abundances (in the usual scale log{\it N}(X) $+$ 12) are indicated. Synthetic
spectra obtained for Li and Rb abundances shifted $+$0.5 dex (in blue) and -0.5
dex (in green) (these values are $\pm$1.0 for Li in R Cen and SV Cas) from the
adopted values are also shown. We note that Li is not detected in the extreme
{\it OH/IR} AGB star RU Ari, which displays a strong Rb I line that is not
detected in the other stars.} \label{FigA1}
\end{figure*}

\begin{figure*}
\centering
\includegraphics[angle=0,scale=.35]{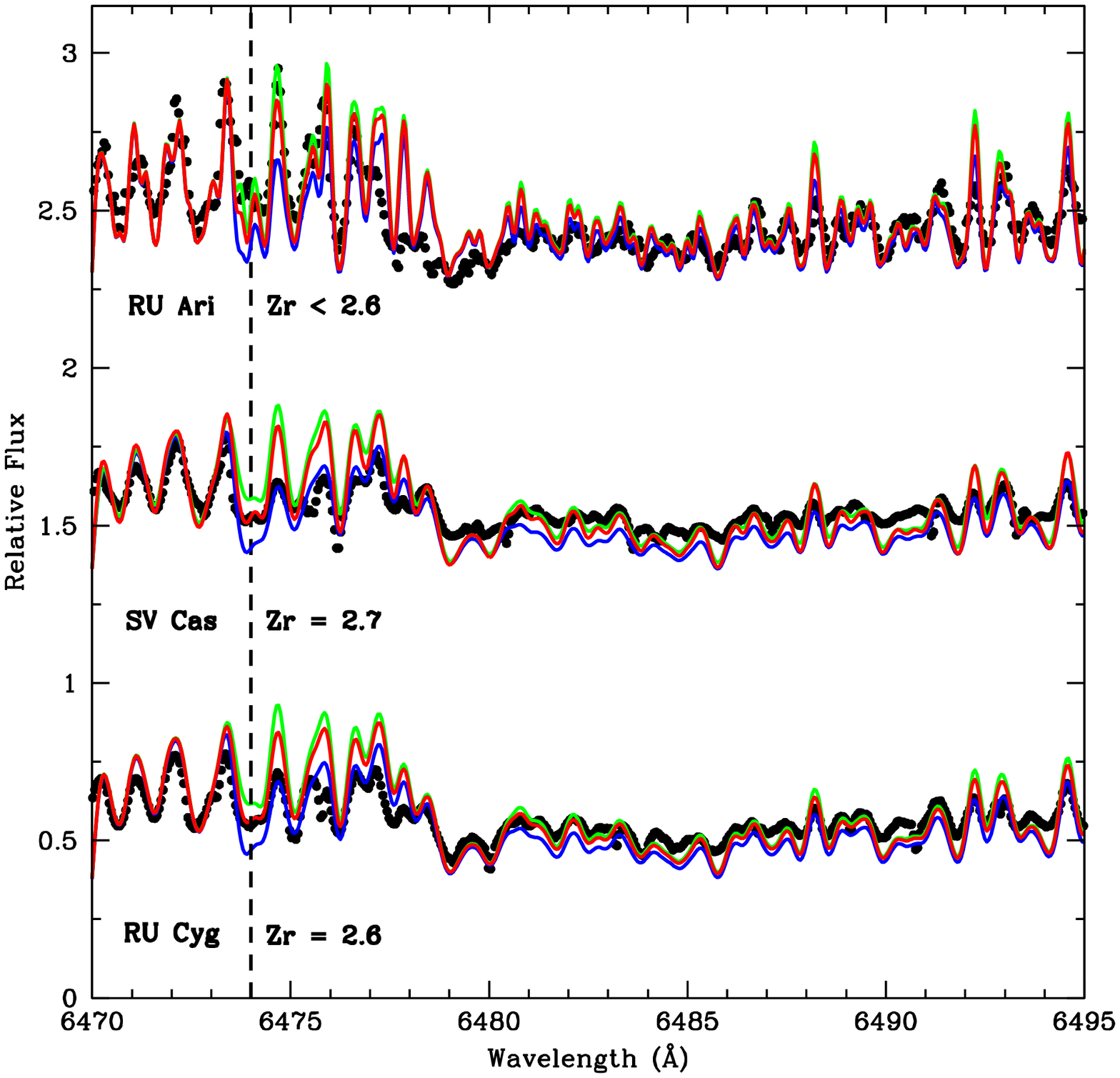}%
\includegraphics[angle=0,scale=.35]{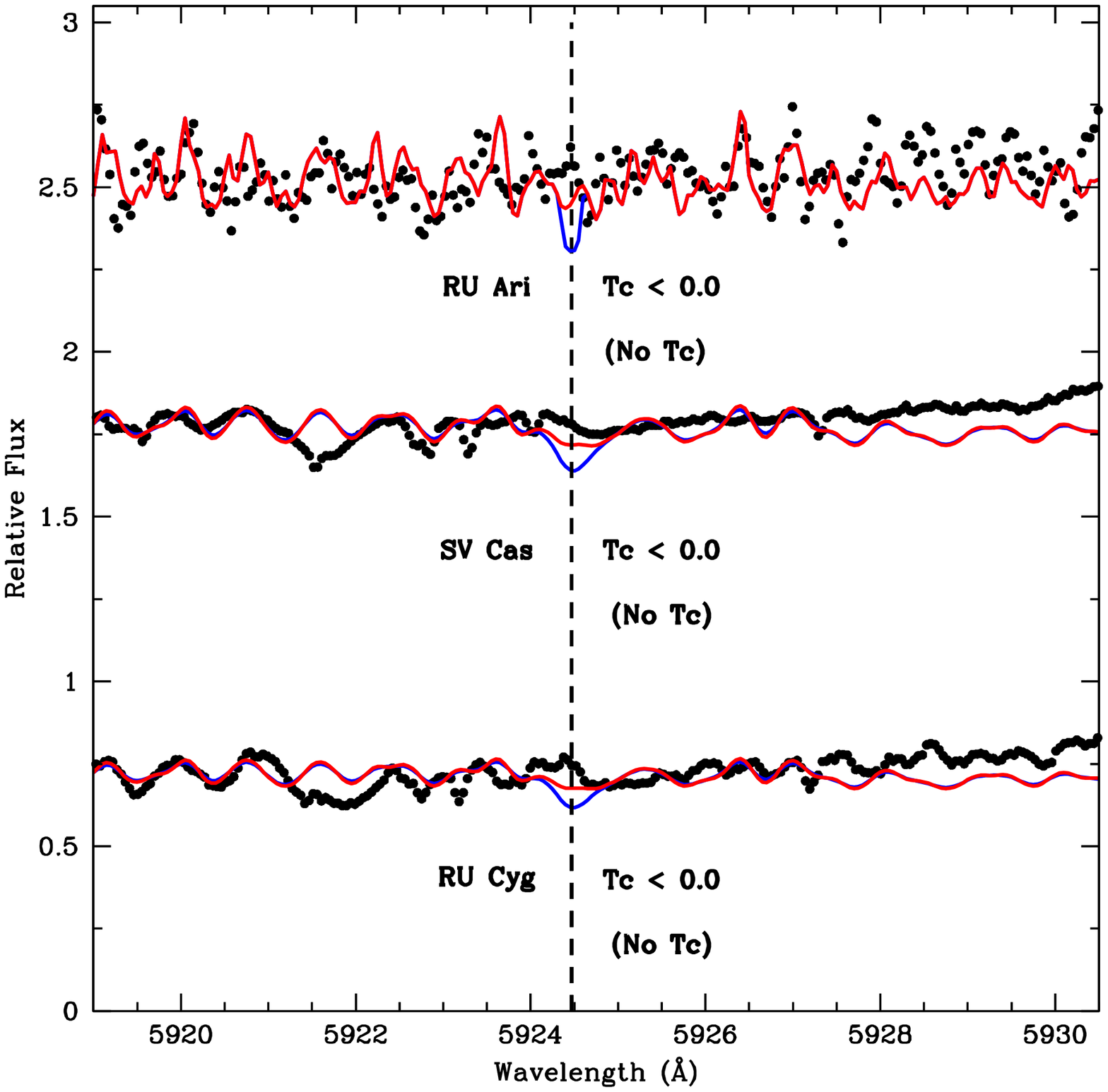}
\caption{High-resolution optical spectra (in black) and best model fits (in red)
in the ZrO 6474\AA~region (left panel) and the Tc I 5924\AA~region (right panel)
for the AGB stars RU Ari, SV Cas, and RU Cyg. The derived Zr abundances and Tc
upper limits (in the usual scale log{\it N}(X) $+$ 12) are indicated. Synthetic
spectra obtained for Zr abundances shifted $+$0.5 dex (in blue) and -0.5 dex (in
green) from the adopted values are also shown. For the less sensitive Tc I
5924\AA~line, synthetic spectra obtained for Tc abundances shifted $+$5.0 dex
(in blue) are shown. We note that none of the stars are in our sample is found
to be enriched in Zr or Tc.}
\label{FigA2}%
 \end{figure*}

\begin{figure*}
\centering   
\includegraphics[angle=0,scale=.35]{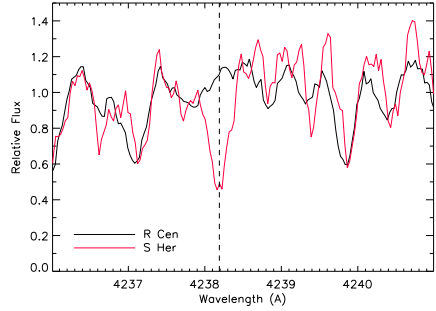}
\caption{High-resolution optical spectra of the non-Tc AGB star R Cen (in black;
this paper) and the Tc AGB star S Her (in red; taken from Uttenthaler et al.
2011) around the Tc I line at 4238 \AA. The wavelength of the Tc line is marked
with a dashed vertical line. We note that S Her displays a strong Tc I line that
is completely absent in R Cen.} \label{FigA3}
\end{figure*}

\begin{figure*}
\centering   
\includegraphics[angle=90,scale=.35]{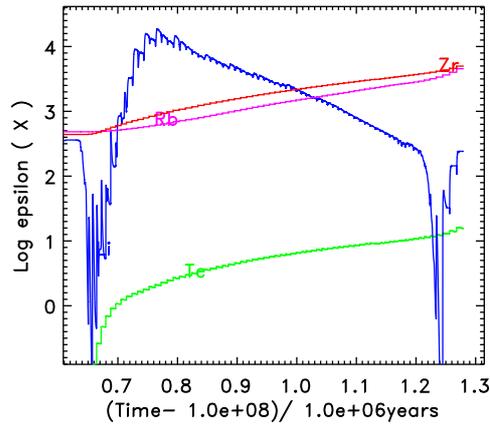}
\caption{Temporal evolution of the Li, Rb, Zr, and Tc abundances as predicted by
the HBB-MLT 5 M$_{\odot}$ model with a $^{13}$C-rich region (the $^{13}$C
pocket) and a delayed superwind (Karakas et al. 2012). We include a $^{13}$C
pocket by inserting protons into the top of the He-intershell (with a mass $= 1
\times 10^{-4} M_{\odot}$) at the deepest extent of each TDU episode. We refer
to Karakas et al. (2012) and Lugaro et al. (2012) for details of this procedure.
Tc (also Zr although at a slower rate) is quickly produced by the $^{13}$C
neutron source; log$\varepsilon$(Tc)$>$0 at the beginning of the super Li-rich
phase (log$\varepsilon$(Li)$\sim$4). The evolution of Rb is not greatly affected
by the inclusion of the $^{13}$C pocket, but Zr is more affected, being even
more abundant than Rb during the  super Li-rich phase. We note that the Tc
abundance is an upper limit because the trend of the half-life of $^{99}$Tc, 
decreasing with the temperature, is not included. However, the difference would
be very small because the $^{99}$Tc half-life changes only from terrestrial 0.22
Myr to 0.11 Myr at the temperature of 100 MK typical of the $^{13}$C pocket,
which means that $^{99}$Tc behaves as a stable nucleus during this neutron
flux.}
\label{FigA4}
\end{figure*}

\end{appendix}

\end{document}